\documentclass[%
superscriptaddress,
 amsmath,amssymb,
 aps,
twocolumn
]{revtex4-2}

\usepackage{graphicx}
\usepackage{dcolumn}
\usepackage{bm}
\usepackage{xcolor,soul}

\usepackage{hyperref}

\usepackage[normalem]{ulem}

\let\a=\alpha \let\b=\beta \let\g=\gamma \let\d=\delta
\let\e=\epsilon \let\z=\zeta  \let\k=\kappa
\let\l=\lambda \let\m=\mu   \let\p=\pi
\let\s=\sigma  \let\f=\varphi 
   \let\G=\Gamma
\let\D=\Delta \let\Th=\Theta  
   
 \let\r=\rho  \let\io=\infty

\def\ie{{i.e. }}

\def\ZZ{{\cal Z}}

\def\de{\mathrm d}

\def\tk{\widetilde{\kappa}}

\def\to{\rightarrow}

\newcommand{\beq}{\begin{equation}} \newcommand{\eeq}{\end{equation}}
 
\newcommand{\argc}[1]{\left[#1\right]}
\newcommand{\arga}[1]{\left\lbrace #1\right\rbrace }
\newcommand{\argp}[1]{\left(#1\right)}

\newcommand{\moy}[1]{\left\langle  #1 \right\rangle }

\DeclareRobustCommand{\vect}[1]{
  \ifcat#1\relax
    \boldsymbol{#1}
  \else
    \mathbf{#1}
  \fi}

  \newcommand{\av}[1]{\left\langle#1\right\rangle}
  \newcommand{\cbr}[1]{\left(#1\right)}
  \newcommand{\sbr}[1]{\left[#1\right]}

  \newcommand{\new}[1]{#1}
  \newcommand{\old}[1]{}
  \newcommand{\newAM}[1]{#1}
\newcommand{\blue}[1]{#1}
\newcommand{\red}[1]{#1}

\begin{document}

\title{Control protocols for harmonically confined run-and-tumble particles}

\author{Marco Baldovin}
 \email{marco.baldovin@cnr.it}
\affiliation{Institute for Complex Systems, CNR, 00185, Rome, Italy}

\author{Alessandro Manacorda}
\affiliation{Institute for Complex Systems, CNR, 00185, Rome, Italy}

\date{\today}

\begin{abstract}
Run-and-tumble particles constitute one of the simplest models of self-propelled active matter, and provide an ideal playground to the understanding of out-of-equilibrium systems. We consider an idealized setup where  one such particle is subject to a harmonic confining potential, and an external agent can vary in time the tumbling rate and the strength of the trap. We search for time-dependent control protocols steering the system between assigned end states, in a prescribed time interval. To this aim, we propose a description of the dynamics, alternative to the usual ones, in the form of an infinite set of ordinary differential equations. Solutions based on a suitable closure of such hierarchy, which we expect to hold true in the limit of long protocol duration, are discussed and compared with numerical simulations. We also look for the protocol completing the task with the minimal work, on average: the problem can be tackled analytically, again in the regime of slow (but not quasi-static) transformations. The solution provides insightful intuition on the optimal strategies for the control of active matter systems.

\end{abstract}

\maketitle


\section{Introduction}

Active matter systems are characterized by the ability to sustain nonequilibrium steady states through the continuous consumption of energy at the microscopic scale. Unlike passive systems in contact with a thermal bath, their stationary distributions generally violate detailed balance and cannot be expressed in Boltzmann form. This fundamental property underlies many of the distinctive behaviors of active systems, ranging from persistent motion to collective organization, and has been extensively discussed in the literature on active matter physics~\cite{Marchetti13rmp,Bechinger16rmp}.

One of the most striking consequences of nonequilibrium steady states is the possibility of extracting useful work from active systems even in the presence of a single thermal reservoir. Active engines and ratchet mechanisms have been proposed and analyzed as paradigmatic examples of this phenomenon~\cite{DiLeonardo10pnas,Pietzonka19prx,Ekeh20pre, gronchi2021}, highlighting how activity can be converted into mechanical work without violating the second law of thermodynamics~\cite{Mandal17prl,Pietzonka18prl}. These results have motivated a growing interest in the energetic performance of active systems and in their potential use as microscopic work-extracting devices.

In this context, the problem of control naturally arises. \blue{A distinctive feature of many experimental realizations of active matter is the possibility of controlling particle dynamics through external fields, most notably light. In particular, light-responsive systems such as synthetic active colloids~\cite{Buttinoni12jpcm}, micromotors~\cite{Maggi15natcomm} and photokinetic bacteria~\cite{Vizsnyiczai17natcomm} allow for a direct tuning of propulsion speed, which in turn provides a handle on the spatial distribution of particles. This mechanism has been successfully exploited to engineer nontrivial stationary density profiles and even spatial patterns~\cite{Frangipane18elife}.}

For passive stochastic systems, it is well established that performing a transformation slowly is generally sufficient to minimize dissipation, recovering the quasi-static limit of equilibrium thermodynamics. By contrast, in active systems, slowness alone does not guarantee minimal dissipation~\cite{Davis24prx}. The intrinsic nonequilibrium nature of steady states implies that even infinitely slow protocols may incur a finite energetic cost, and the outcome of a transformation depends in a nontrivial way on the chosen driving protocol. This observation calls for a systematic theory of optimal control beyond the quasi-static paradigm.

The issue becomes particularly acute when considering transformations performed in a finite time. In passive systems, the framework of shortcuts to adiabaticity has shown that suitably designed control protocols can reproduce adiabatic transformations within a finite duration, often leading to general relations between dissipation, protocol duration, and control parameters~\cite{GueryOdelin19rmp}. Universal features of optimal finite-time driving have been found in equilibrium and near-equilibrium settings~\cite{Schmiedl07prl,Aurell11prl}. 
Recent studies have extended the focus to out-of-equilibrium systems, including the so-called Brownian Gyrator~\cite{baldassarri2020} and other off-equilibrium linear models~\cite{lucente2025optimal}, stochastic dynamics with non-Gaussian~\cite{baldovin2022} and 
non-Markovian noise~\cite{loos2024universal}, stochastic resetting~\cite{de2023resetting,goerlich2024resetting,goerlich2026time} and granular materials~\cite{prados2021, ruiz2022optimal}.  However, the situation is much more challenging in these cases: steady states are non-Boltzmann, detailed balance is broken, and no general adiabatic reference transformation exists. As a result, no universal relations for shortcuts and optimization have been identified so far.


Several attempts have been made to address optimal control in active systems. Some approaches are based on a formulation of the control problem directly at the level of probability distributions~\cite{Baldovin23prl, Frim23arxiv}. From this perspective, the task of driving a system from an initial to a final steady state can be seen as an optimal transport problem in the space of distributions, where the full structure of the nonequilibrium steady states is explicitly taken into account.
The drawback of this approach is that it is analytically feasible only for a few models, as it requires explicit knowledge of the functional form of the distribution. Another strategy is based on the   formulation of the problem in the linear-response regime~\cite{Davis24prx}, assuming that the protocol duration is long with respect to the characteristic times of the dynamics. Alternatively, one can focus on the evolution of a limited set of observables, such as the first few moments of the distribution\blue{~\cite{Schuttler25pre,GarciaMillan25prl,olsen2025harnessing}}, in order to get exact, insightful results.

In this work, we focus on the dynamics of a run-and-tumble particle (RTP) in one spatial dimension: the model is simple enough to be analytically studied, but it shows nontrivial properties that are typical of active matter~\cite{Schnitzer93pre,PhysRevLett.100.218103,Solon15natphys,malakar2018steady,paoluzzi2024entropy,paoluzzi2025local}.

We address the problem of control, taking inspiration from all the approaches described above. We first determine a set of equations for the time-dependent control parameters that transport the system between prescribed nonequilibrium steady states, with a reverse engineering strategy. The framework is similar to a cumulant expansion, and it leads to an infinite hierarchy of equations. In order to obtain analytical results, we then focus on the limit of long protocol duration and, assuming to be in the linear response regime, we make suitable assumptions to truncate the hierarchy.
Within this setting, we show that optimal protocols can be obtained analytically at leading order in the perturbative expansion, providing explicit expressions for the control parameters and the associated energetic cost. Our results suggest a systematic route to the construction of optimal protocols in active matter and, more broadly, in a wide class of nonequilibrium stochastic systems.

The paper is organized as follows. In Section~\ref{sec:model} we describe the RTP model in one dimension, in the absence of thermal noise. In Section~\ref{sec:equiv} we reformulate the dynamics as an infinite hierarchy of ordinary differential equations, with a method that resembles cumulant expansion. Section~\ref{sec:short} is devoted to the study of shortcuts for the system: explicit results are obtained in the limit of long protocol duration. The optimization of such protocols is discussed in Section~\ref{sec:opti}. Finally in Section~\ref{sec:conc} we draw our conclusions.

\section{Model}
\label{sec:model}
    We consider the dynamics of a RTP in one dimension. Its state is determined by its position $x$ and its orientation $\s=\pm1$. The particle's state evolves according to
    \beq\label{eq:micro1}
        \dot x(t) = - \m U'(x(t)) + v_0 \s(t) \ ,
    \eeq
    being the orientation $\s(t)$ a random telegraph process with tumbling rate $\a/2$, \ie correlation time $1/\a$, and $v_0$ the particle's self-propulsion. \newAM{Furthermore, the particle is confined in space by an external potential $U(x)$ and the dynamics is overdamped with a mobility coefficient $\m$.}

    The probability density of finding a particle at $x$ with positive (negative) orientation is $P_+(x)$ ($P_-(x)$). It is useful to define the density $\r(x)$ and magnetization $m(x)$ as
    \beq\label{eq:rho-m}
        \r(x) = P_+(x) + P_-(x) \ , \quad m(x) = P_+(x) - P_-(x) \ .
    \eeq
    \old{Furthermore, the particle is confined by an external potential $U(x)$ and the dynamics is overdamped with a mobility coefficient $\m$.
    In the continuous time limit $\D t\to 0$, the} \newAM{The} density and magnetization \newAM{fields} \blue{ evolution in the continuous time limit can be computed~\cite{Schnitzer93pre,Solon15natphys}, starting from the evolution of the probability densities $P_\pm(x,t)$}
    \begin{subequations}\label{eq:dyn-pm}
        \begin{eqnarray} 
            \nonumber \partial_t P_+(x,t) &=& - \partial_x \left[ v_0 P_+(x,t) + \m U'(x) P_+(x,t) \right] \\
            & &- \frac{\a}2 \left[ P_+(x,t) - P_-(x,t) \right] \ , \\ 
            \nonumber \partial_t P_-(x,t) &=&  \partial_x \left[ v_0 P_-(x,t) - \m U'(x) P_-(x,t) \right] \\
            & &+ \frac{\a}2 \left[ P_+(x,t) - P_-(x,t) \right] \ , \label{eq:dyn-pm2}
        \end{eqnarray}
    \end{subequations}
    \blue{leading to the Chapman-Kolmogorov equations}
    \begin{subequations}\label{eq:dyn}
        \begin{eqnarray} \label{eq:dyn1}
            \partial_t \r(x,t) &=& - \partial_x \left[ v_0 m(x,t) - \m U'(x) \r(x,t) \right] \, , \\ \nonumber \partial_t m(x,t) &=& - \partial_x \left[ v_0 \r(x,t) - \m U'(x) m(x,t) \right]\\
           & &
           \label{eq:dyn2}  - \a m(x) \ .
        \end{eqnarray}
    \end{subequations} 
    We highlight the absence of thermal noise in this framework, \ie $T=0$. Therefore, the only source of fluctuations is the particle's activity. If the speed $v_0$ and the mobility $\m$ are non-vanishing, we can set a time and length scale assuming $v_0=\m=1$ and working in dimensionless units. The active system admits a steady-state solution that can be computed exactly and, as expected, differs from the Boltzmann distribution~\cite{Tailleur09epl}. In the general case, one has
    \begin{subequations}
        \begin{eqnarray}
            \r_s(x) &=& \frac{C_0}{1-(U'(x))^2} \exp \argc{ -\a \int \frac{U'(x)}{1- (U'(x))^2} \de x} \ , \\
            m_s(x) &=& U'(x) \r(x) \ .    
        \end{eqnarray}
    \end{subequations}
            
    We will focus on the harmonic potential case $U(x) = \dfrac12 \k x^2$ from now on. In this case, the stationary density reads
    \begin{subequations}\label{eq:rhos-harmonic}
    \begin{eqnarray}
        \r_s(x) &=& \frac{\k}{\ZZ_s(\b)} (1 - \k^2x^2 )^\b \Th \left( 1 - \k^2 x^2 \right) \ , \\ 
        \ZZ_s(\b) &=& \sqrt\p \frac{\G\left(\b+1\right)}{\G\left(\b+\frac32\right)} \ . \label{eq:rhos-harmonic-ZZ}
    \end{eqnarray}
    \end{subequations}
    Here, $\Th(x)$ represents the Heaviside theta function and $\b = \a/(2\k)-1$. The density is then given by a Beta distribution, defined in the finite domain $\vert x \vert < 1/\k$ and governed by the $\b$ exponent. Thus, the system displays a transition between a passive-like state with $\b>0$, where the density is peaked at $x=0$ and tends to a Gaussian when $\a\to\io$, to an active state where the density accumulates at the boundaries $x = \pm 1/\k$ when $\b<0$. The normalization coefficient linearly depends on the stiffness $\k$ for dimensional reasons, and its main complexity is enclosed in the $\ZZ(\b)$ factor defined in Eq.~\eqref{eq:rhos-harmonic-ZZ}. 

\begin{figure}
        \centering
        \includegraphics[width=0.99\linewidth]{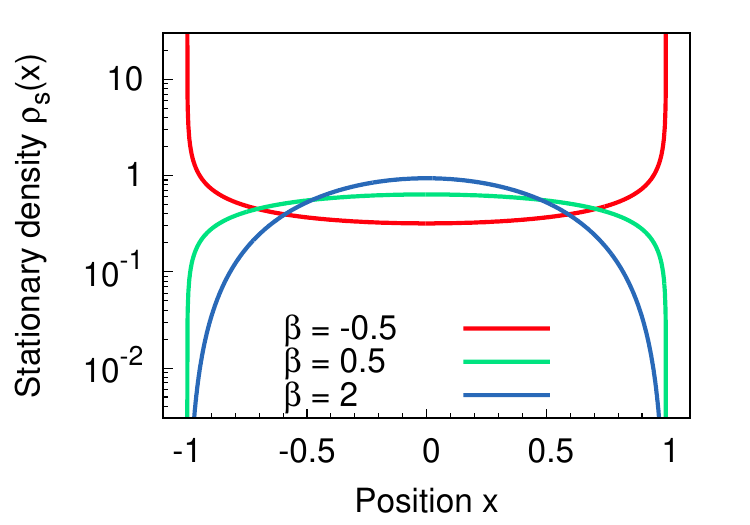}
        \caption{Stationary distribution of the RTP process in the presence of harmonic confinement, Eq.~\eqref{eq:rhos-harmonic}. \newAM{The red line corresponds to the active case ($\b<0$), while the green and blue lines to the passive case ($\b>0$)}. \old{Different values of $\beta$ are considered, while} \newAM{The stiffness is set to $\k=1$.}}
        \label{fig:activepassive}
    \end{figure}

    Physically speaking, the active state arises when the tumbling rate $\a$ is so low that particles remain in the vicinity of the frontier $\vert x \vert = 1/\k$, where self-propulsion and the external harmonic force are perfectly balanced. In this case, only a few particles are able to reverse their motion and escape the frontier towards the bulk of the system. Conversely, for high tumbling rate $\a$ or low stiffness $\k$, the particles frequently tumble and never get trapped at the frontiers, keeping the bulk highly populated. The situation is represented in Fig.~\ref{fig:activepassive}, where the stationary position distribution is shown for different values of $\beta$. The stiffness and the tumbling rate then completely determine the system state, and will represent our control parameters from now on.

\blue{In the presence of thermal noise ($T>0$)  the dynamics is described by a Fokker-Planck equation. It turns out that exact shortcuts can be found~\cite{Frim23arxiv} by manipulating in time the parameters of the pdf. In the present case, which can be seen as a singular limit of the diffusive model, that approach is not feasible, and we need to rely on a suitable reformulation of the problem. }

\section{An equivalent formulation of the dynamics}
\label{sec:equiv}



    We aim to find the time-dependent controls $(\k(t),\a(t))$ capable of steering the density and magnetization from a given initial state to a given final state \textit{in a finite time} $t_f$, \ie $(\r_i(x),m_i(x)) \;\xrightarrow[t_f]{}\;(\r_f(x),m_f(x))$.

    The solution to this problem typically requires the knowledge of the state evolution as a function of the control parameters, which is generally a  challenging task for non-Gaussian dynamics with time-dependent coefficients.
    So far, an explicit analytical solution for the dynamics of the distribution is missing. We therefore propose an alternative route to answer this question: the time-dependent density can be written as
    \beq\label{eq:gc5}
        \begin{aligned}
            \r(x,t) =& \frac{\tk(t)}{\mathcal{Z}(\vect{\b}(t))}\exp\argp{-\sum_{n=1}^{\infty}\frac{\b_n(t) \tk^{2n}(t)}{n} x^{2n}} \\
            & \times \Theta\left[1-\tk^2(t)x^2\right] \ .
        \end{aligned}
    \eeq
    Here, we introduced the time-dependent \textit{distribution parameters} $\tk(t)$ and $\vect{\b}(t)=(\b_1(t),\b_2(t),\ldots)$. The former represents the reciprocal of the frontier position at time $t$, \ie the density $\r(x,t)$ vanishes for $\vert x \vert > 1/\tk(t)$. The latter is an infinite-dimensional vector of coefficients determining the density moments. \red{This representation shifts the problem from solving a partial differential equation for the full probability density to a dynamical system for a set of scalar observables. This is advantageous because it allows for a direct construction of control protocols, even when the full time-dependent distribution is not analytically accessible}.
    The normalization coefficient at time $t$ reads
    \beq\label{eq:N}
        \ZZ(\vect{\b}) = \int_{-1}^{1} \,du\, \exp\argp{-\sum_{n=1}^{\infty}\frac{\b_n}{n} u^{2n}} \, . 
    \eeq
     \new{Since the center of the well does not vary in time, the density is \old{always} even} \newAM{at all times} \new{and therefore only the even powers of $x$ appear in the series. The above expression is, in this respect, totally general. In the stationary case all the $\{\b_n\}_{n\ge1}$ are equal, and a Beta distribution is recovered.}
     
     \new{It is important to notice that the series appearing in Eq.~\eqref{eq:gc5} cannot be truncated, because it appears as the argument of an exponential. The expansion will be nonetheless useful to carry on analytical calculations. If a slow dynamics is considered, we will assume that the distribution is close to the stationary one, and all $\{\b_n\}$ are equal for $n$ larger than a certain threshold (see Section~\ref{sec:qs} below).}
     
    We then write the magnetization as
    \beq\label{eq:gc3a}
        m(x,t)\equiv \g(x,t)\r(x,t)  x \,,
    \eeq
    where $\g(x,t)$ is an even function of its argument. Conversely to the density, the magnetization is an odd field for isotropic systems and therefore only odd powers of $x$ appear in its power expansion. By introducing a third set of parameters $\vect{c}(t)=(c_1(t),c_2(t),\ldots)$, we can expand such function as
    \beq\label{eq:gc3}
        \g(x,t)=  \sum_{n=0}^{\infty}c_n(t) x^{2n}\,.
    \eeq
    We generally omit the time-dependence of controls and parameters from now on. The stationary state at $(\k,\a)$ corresponds to the parameters

    \beq\label{eq:ss-parameters}
        \tk=\k \ , \quad \b_n = \b = \a/(2\k)-1 \ , \quad c_n=\k \,\d_{n,0} \ .
    \eeq

    The ansatz in Eqs.~\eqref{eq:gc5}-\eqref{eq:gc3} is the key assumption of our work: we map the problem of determining the evolution of two fields $\r$ and $m$ to the evolution of a set of state parameters $\tk$, $\{\b_n\}_{n\ge1}$ and $\{c_n\}_{n\ge1}$. The goal is now to determine their evolution as functions of the control parameters $(\k,\a)$ and to solve the inverse problem to find the corresponding control.

    To perform the mapping of the dynamics of Eq.~\eqref{eq:dyn} into the parameters dynamics, it is useful to introduce the following quantities
    
    \begin{subequations}\label{eq:gc}
        \begin{eqnarray}
            \label{eq:w} w&\equiv&\partial_t \ln \mathcal{Z}\,,\\
            \label{eq:phi}   \f_n&\equiv& \frac{\dot{\b_{n}}}{n}+2\b_n w \,,\\
            \label{eq:zeta} \z_n&\equiv&\b_n\tk^{2}-\b_{n-1} c_0^2+\f_{n-1}c_0\,.
        \end{eqnarray}
    \end{subequations}

    After some algebra   one finds the following relations:
        \begin{subequations}\label{eq:gc}
            \begin{eqnarray}
                \label{eq:gc_f5}\k &=& \frac{\dot{\tk}}{\tk}+ \tk\,,\\
                \label{eq:gc_f3}\a &=& \k +2\frac{\b_1\tk^{2}}{c_0}+c_0-\frac{\dot{c}_0}{c_0}\,,\\
                \label{eq:gc_f1}c_0&=&\tk+w\,,\\
                \label{eq:gc_f2}\argp{2n+1}c_n&=&2\sum_{j=1}^{n-1}c_j\b_{n-j}\tk^{2\argp{n-j}}+\tk^{2n}\f_n\quad\\
                \label{eq:gc_f4}2\argp{2n+1}\z_{n+1}&=&\sum_{j=1}^{n-1}\argp{4\b_{n-j}\z_{j+1}-\f_j \f_{n-j}}\\
                \nonumber& &+\dot{\f}_n+\argp{2n w +\a-\k}\f_n \,,
            \end{eqnarray}
        \end{subequations} 
        where the last two equations hold for every $n\ge1$.
    The strategy to obtain this result is detailed in Appendix~\ref{app:comp}.
    \new{Let us notice that, with our parametrization, the coefficients $\{c_n(t)\}_{n\ge 0}$ are completely determined once the parameters $\tk(t)$ and $\{\b_n\}_{n\ge1}$ of the density are known.}

\section{Shortcuts}
\label{sec:short}
The new formulation~\eqref{eq:gc} of the dynamics of the system provides us with a powerful tool to compute \textit{shortcuts} between assigned states. This is a classical problem in quantum mechanics (in that context, one speaks of ``shortcuts to adiabaticity''),  which has been recently extended also to stochastic thermodynamics~\cite{guery2023driving}. 
\red{The strategy is as follows. First, we parametrize the state of the system in terms of $\tk$ and $\{\beta_n\}$. Second, we derive their dynamics as functions of the control parameters. Finally, we invert these relations to construct protocols that connect assigned initial and final states.}

Suppose that density and \blue{magnetization} are known at time $t=0$, i.e.
\begin{equation}
\nonumber
    \begin{aligned}
        \r(x,0)&=\r_i(x)\\
        m(x,0)&=m_i(x)\,.
    \end{aligned}
\end{equation}
We wish to steer the system in such a way that,  at a definite time $t=t_f$, it reaches a target state characterized by 
\begin{equation}
\nonumber
    \begin{aligned}
\r(x,t_f)&=\r_f(x)\\
m(x,t_f)&=m_f(x)\,.
    \end{aligned}
\end{equation}

\blue{We can modify the state of the system by acting on the parameters $\k(t)$ and $\a(t)$, which ideally represent the control parameters of the experiment.}
In our description, the state of the system at a certain time is \blue{instead} determined by the values of $\tk$ and $\{\b_n\}_{n\ge 1}$. The problem consists therefore in finding a set of functions $\tk(t)$, $\{\b_n(t)\}_{n\ge1}$ fulfilling Eqs.~\eqref{eq:gc} and compatible with the boundary conditions at time $t=0$ and $t=t_f$\blue{, and then determining the controls $\k(t)$ and $\a(t)$ enforcing such evolution}. Since the dynamics involves an infinite number of ODEs, in general it does not appear any easier than directly finding a solution of the original PDEs.

As a first remark, it can be noticed that if $\b_1(t)$, $\tk(t)$ and $w(t)$ were known in the time interval $(0,t_f)$, then Eq.~\eqref{eq:gc_f1} would provide an expression for $c_0(t)$, and the controls $\a(t)$ and $\k(t)$ could be determined by means of Eqs.~\eqref{eq:gc_f5} and~\eqref{eq:gc_f3}. All $\{\b_n(t)\}_{n\ge2}$ could be then obtained, in principle, by applying Eq.~\eqref{eq:gc_f4}. At that point, however, the system would be overspecified, because of Eq.~\eqref{eq:w} relating $w(t)$ to the  $\{\b_n(t)\}$. 
This suggests a possible strategy to build shortcuts.

\blue{We detail in Appendix~\ref{sec:rec} a recursive strategy to reach this target in the general case, which is however prone to technical difficulties. In what follows we limit ourselves to the case of long protocol duration, which allows us to write down analytical expressions.}

\subsection{Long protocol duration}
\label{sec:qs}

\begin{figure*}[ht]
    \centering
    \includegraphics[width=.8\linewidth]{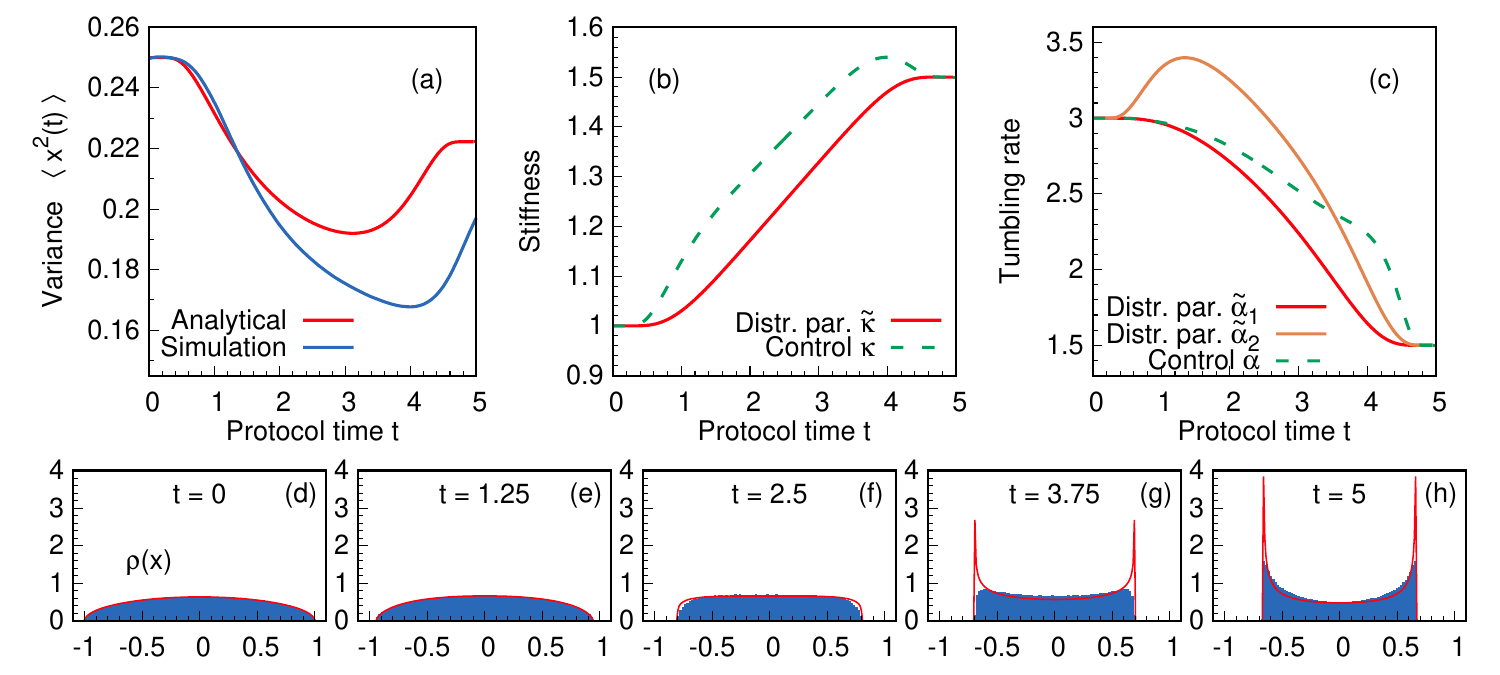}
    \caption{Shortcut in the long-duration regime. In panel (a) the analytical prediction (red) and numerical simulations (blue) for the variance are compared. Panel (b) shows the assigned distribution parameter $\tk(t)$ (solid red) defined by Eqs.~\eqref{eq:simuk} and~\eqref{eq:gf},  and the  control $\k(t)$ (dashed green), analytically computed in Eq.~\eqref{eq:kappapert}. Panel (c) displays the  distribution parameter $\widetilde{\a}_1(t)$ (solid red), defined by Eq.~\eqref{eq:a1}, where $\b_1(t)$ is given by Eqs.~\eqref{eq:simub} and~\eqref{eq:gf}. The parameter $\widetilde{\a}_2$ (solid yellow), computed analytically by means of Eqs.~\eqref{eq:a2} and~\eqref{eq:closure2},  and the control $\a$ (dashed green) defined by Eq.~\eqref{eq:alphapert} are also shown. Finally, in panels (d)-(h) the instantaneous distribution of the process is plotted at different times, both for the analytical prediction~\eqref{eq:rhopert} (red solid line) and the numerical simulation (blue histogram). Parameters: $t_f=5$, $\tk_i=1$, $\tk_f=1.5$, $\b_i=0.5$, $\b_f=-0.5$.}
    \label{fig:nonopt1}
\end{figure*}

\begin{figure*}
    \centering
    \includegraphics[width=.8\linewidth]{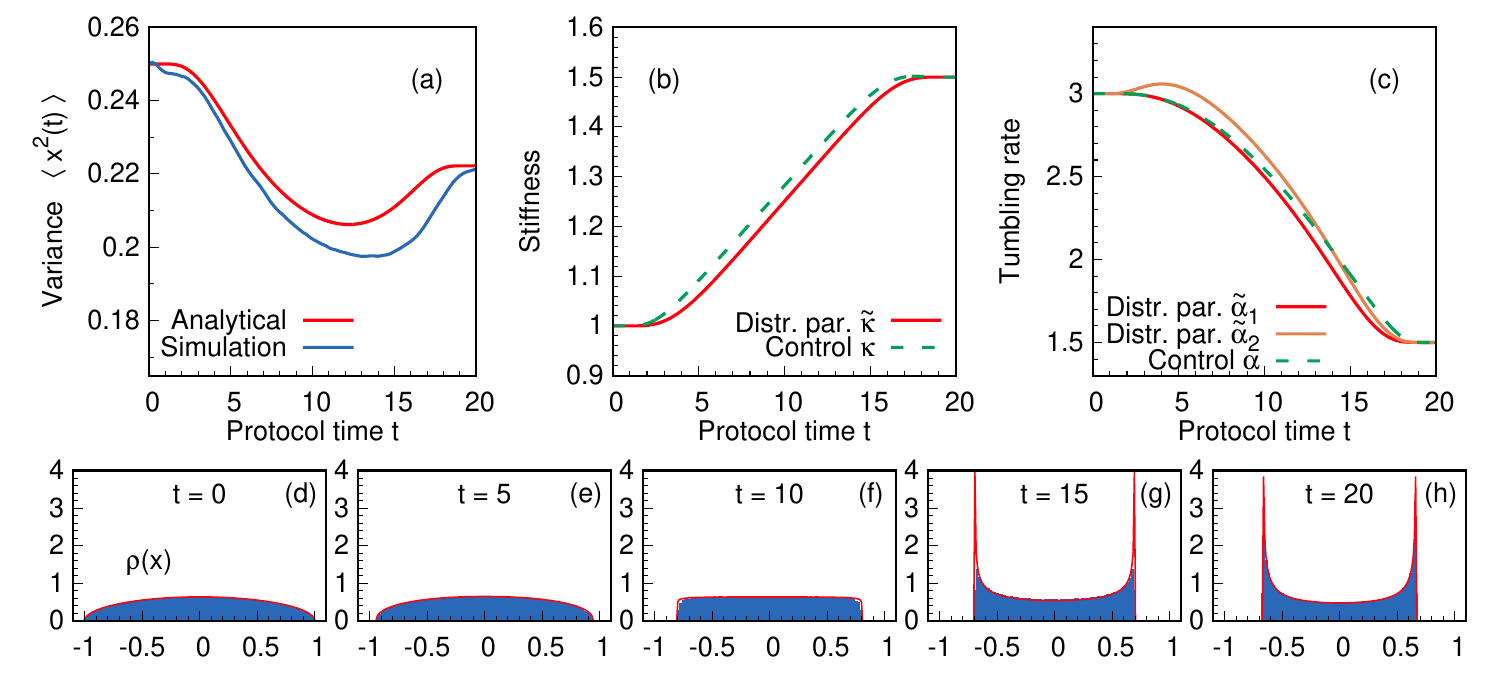}
    \caption{Shortcut in the \new{long-duration} regime. Same as in Fig.~\ref{fig:nonopt1}, with $t_f=20$.}
    \label{fig:nonopt3}
\end{figure*}

Let us assume that the duration $t_f$ of the protocol is much longer than the typical time scales of the system, which we suppose to be $O(\e t_f)$, with $\e \ll 1$. By introducing the rescaled time $s=\e t$, and assuming that the evolution of the state of the system happens on that time scale, one has from Eq.~\eqref{eq:w}:
\begin{equation}
    \label{eq:wpert} w=\e \,\partial_s \ln \mathcal{Z}\equiv \e\widetilde{w}\,.
\end{equation}
Taking into account Eqs.~\eqref{eq:gc_f5}-\eqref{eq:gc_f4}, and noticing that $w=O(\e)$, the controls can be written as:
\begin{subequations}
\begin{align}
\label{eq:kappapert} \k &= \tk + \e \frac{\tk'}{\tk}\,\\
   \label{eq:alphapert} \a &= 2\tk(1+\b_1) + \e \widetilde{w}  - 2 \b_1  \e \widetilde{w}  + O(\e^2)    \,.  
\end{align}
\end{subequations}
Here and in the following, we denote with the prime symbol the derivative with respect to $s$.
In order to express $\a$ in terms of the given state functions $\b_1$ and $\tk$, we need the explicit form of $\widetilde{w}$ or, equivalently, $\mathcal{Z}$. In principle, this can only be obtained by solving the infinite hierarchy of equations~\eqref{eq:gc}. In practice, since we are interested  in the limit of long protocol duration, we expect the distribution to be close to the stationary one, where all $\{\b_n\}_{n \ge 1}$ are equal: it is therefore reasonable to enforce a truncation that takes into account this \textit{a priori} knowledge of the regime we are studying. In particular, we require that
\begin{equation}
    \b_n = \b_2 \quad \forall n >2\,,
\end{equation}
 meaning that $\mathcal{Z}$ effectively depends on two parameters only, namely $\b_1$ and $\b_2$. The truncation therefore preserves the leading non-Gaussian corrections while keeping the problem analytically tractable. This truncation is more realistic than just imposing the stationary form~\eqref{eq:rhos-harmonic-ZZ} at all times of the protocol, and it leads to equations that are still simple enough to be treated analytically, at least to some extent. In particular, one easily obtains from Eq.~\eqref{eq:N}:
 \begin{equation}
 \label{eq:Zpert}
     \mathcal{Z}=\int_{-1}^1 du \, e^{-(\b_1-\b_2)u^2}(1-u^2)^{\b_2}\,.
 \end{equation}

We are left with the problem of expressing $\b_2$ as a function of $\b_1$ and $\tk$. 
First, recalling~\eqref{eq:phi} and~\eqref{eq:zeta}, we notice that
\begin{subequations}
\begin{eqnarray}
\label{eq:phipert}   \f_n&=& \e \argp{\frac{\b_{n}'}{n}+2  \b_n \widetilde{w} }\\
\label{eq:zetapert} \z_n&=&\tk^{2}(\b_n-\b_{n-1})-2\e \b_{n-1} \tk \widetilde{w} +\f_{n-1}c_0\,.
\end{eqnarray}    
\end{subequations}
Equation~\eqref{eq:gc_f4} for $n=1$ leads then to
\begin{equation}
\label{eq:closure2}
     \b_2  \simeq  \b_1 + \frac{\e}{6 \tk}\argp{\b_1 \widetilde{w} + \b_1{\b}'_1+ 2 \b_1^2 \widetilde{w} -5{\b}'_1 } + O(\e^2)\,.
\end{equation}
We can get rid of the explicit dependence on $\widetilde{w}$ in the $O(\e)$ term by expanding~\eqref{eq:Zpert} and considering only the $O(1)$ contribution, which is in fact $\mathcal{Z}_s(\b_1)$ [see Eq.~\eqref{eq:rhos-harmonic-ZZ}]. This implies
\begin{equation}
\label{eq:wpersol}
    \widetilde{w}\simeq \psi(\b_1+1)-\psi(\b_1+3/2) + O(\e)\,,
\end{equation}
where $\psi(x)$ denotes the digamma function. By plugging Eq.~\eqref{eq:wpersol} in Eq.~\eqref{eq:alphapert} we complete the description of the controls at order $O(\e)$. Inserting it in Eq.~\eqref{eq:closure2} leads instead to a $O(\e)$ expression for $\b_2$, and consequently of the state distribution at the same order:
\begin{equation}
\label{eq:rhopert}
    \r(x)\simeq \frac{\tk }{\mathcal{Z}}e^{-(\b_1-\b_2)\tk^2 x^2}(1-\tk^2 x^2)^{\b_2}\Th(1-\tk^2x^2)\,.
\end{equation}

The findings of this section and the goodness of our assumptions can be tested numerically. The idea is to choose arbitrary functions $\b_1(t)$ and $\tk(t)$, slowly varying on a \blue{protocol time $t_f$} $\gg 1$, and computing the controls $\a(t)$ and $\k(t)$. These can then be used in a numerical simulation of the original process~\eqref{eq:micro1}, and the actual distribution obtained in this way can be compared with the analytical prediction. This analysis is reported in Figs.~\ref{fig:nonopt1} and~\ref{fig:nonopt3}, for different values of $t_f$.
Details on the numerical simulations can be found in Appendix~\ref{app:simul}.

\blue{To ensure that the boundary conditions
\begin{subequations}
\begin{eqnarray}
\tk(0) &=\tk_i\,, \quad\quad \tk(t_f) &=\tk_f\\
\b_1(0) &=\b_i\,, \quad\quad \b_1(t_f) &=\b_f \,.
\end{eqnarray}
\end{subequations}
are satisfied,} we choose state functions of the kind
\begin{subequations}
\label{eq:states}
    \begin{eqnarray}
       \label{eq:simub} \b_1(t)&=&\b_i+\argp{\b_f-\b_i}g(t)\\
       \label{eq:simuk} \tk(t)&=&\tk_i+\argp{\tk_f-\tk_i}h(t)\,,
    \end{eqnarray}
\end{subequations}
with
\begin{equation}
\label{eq:gf}
g(t)=h(t)=\frac{1+\mathrm{tanh}\argc{ \,\mathrm{tan}\argp{\frac{\p t}{t_f}-\frac{\p}{2}}}}{2}\,.    
\end{equation}
\red{The choice of $g(t)$ and $h(t)$ is not unique and these functions are not meant to be optimal; they simply provide smooth interpolations satisfying the required boundary conditions. The latter can be found in Appendix~\ref{sec:rec}. }
We also define the quantities
\begin{subequations}
\begin{eqnarray}
   \label{eq:a1} \widetilde{\a}_1 &=2 \tk (1+\b_1) \\
   \label{eq:a2} \widetilde{\a}_2 &=2 \tk (1+\b_2)\,.
\end{eqnarray}
\end{subequations}
In the stationary state, $\widetilde\a_1=\widetilde\a_2=\a$.

As expected, the agreement between the distribution computed analytically and the numerical realization of the protocol gets better and better as $t_f$ increases.

\section{Optimal protocols}
\label{sec:opti}
Once a method to define shortcuts between stationary states has been devised, it is possible to address questions about the optimality of the process. One may ask, for instance, what is the most efficient way of connecting two given states in a given time, in terms of the mechanical work operated on the system. In other words, one may ask what is the protocol for $\k(t)$ and $\a(t)$ such that the particle passes from state $\r_i(x)$ to $\r_f(x)$ in a time $t_f$, while minimizing the cost
\begin{equation}
    \label{eq:cost}
    \begin{aligned}
    \mathcal{W}&=\int_0^{t_f}\, dt\,\av{\partial_t U(x)}\\
    &=\int_0^{t_f}  dt \, \new{\frac{\dot{\k}(t)}{2}}\int  dx\,  \r(x,t) x^2\,.
    \end{aligned}
\end{equation}
The problem is far from trivial, because the minimization must be performed taking into account the dynamical constraints on the evolution of $\r(x,t)$, which in principle depends on $\tk$ and all the $\{\b_n\}_{n\ge 1}$. The full Lagrangian reads
\begin{equation}  
\label{eq:lag}
\begin{aligned}
\mathcal{L}(\tk,\dot{\tk},\ddot{\tk},\vect{\b},\dot{\vect{\b}}, \ddot{\vect{\b}})&=\argc{\frac{\ddot{\tk}}{\tk}-\argp{\frac{\dot{\tk}}{\tk}}^2+\dot{\tk}}\moy{x^2}\\
&+\sum_{n\ge 1} \lambda_{n}b_n(\tk,\dot{\tk},\vect{\b},\dot{\vect{\b}}, \ddot{\b}_{n})\,,     
\end{aligned} 
\end{equation}
where the expression in square brackets is obtained from~\eqref{eq:gc_f5} by differentiation, $\{\l_n\}_{n \ge 1}$ are time-dependent Lagrange multipliers enforcing the dynamical constraints, and
$$
b_n(\tk,\dot{\tk},\vect{\b},\dot{\vect{\b}}, \ddot{\b}_{n})=0
$$
is a compact notation for Eq.~\eqref{eq:gc_f4}. If one was able to solve the Euler-Lagrange equations associated to $\mathcal{L}$ with boundary conditions~\eqref{eq:boundcond}, then one could plug the explicit expression of $\tk(t)$ and $\{\b_n(t)\}_{n\ge 1}$ obtained in this way in Eqs.~\eqref{eq:gc}, to compute the optimal control protocols $\a(t)$ and $\k(t)$.

While the task of studying the Euler-Lagrange equations for the full Lagrangian $\mathcal{L}$ appears out of reach, in what follows we consider the simpler case of the long protocol duration.

\subsection{Long protocol duration}
\label{sec:optqs}

\begin{figure*}
    \centering
    \includegraphics[width=.8\linewidth]{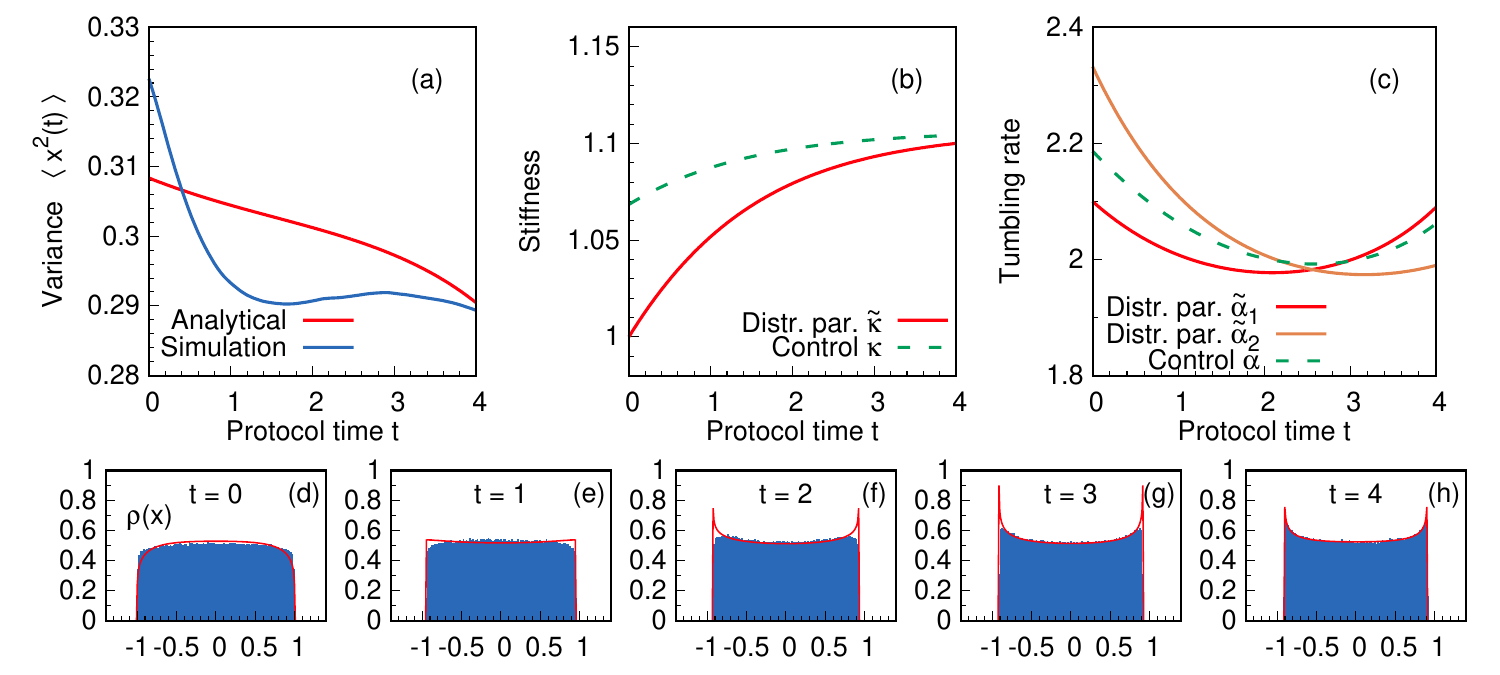}
    \caption{Optimal shortcut in the \new{long-duration} regime. The plotted quantities and the color code are the same as in Fig.~\ref{fig:nonopt1}, but this time the protocol has been obtained by solving (numerically) the Euler-Lagrange equations~\eqref{eq:eulag}. Parameters: $t_f=4$, $\tk_i=1$, $\tk_f=1.1$, $\b_i=0.05$, $\b_f=-0.05$.}
    \label{fig:opt}
\end{figure*}

\begin{figure*}
    \centering
    \includegraphics[width=.8\linewidth]{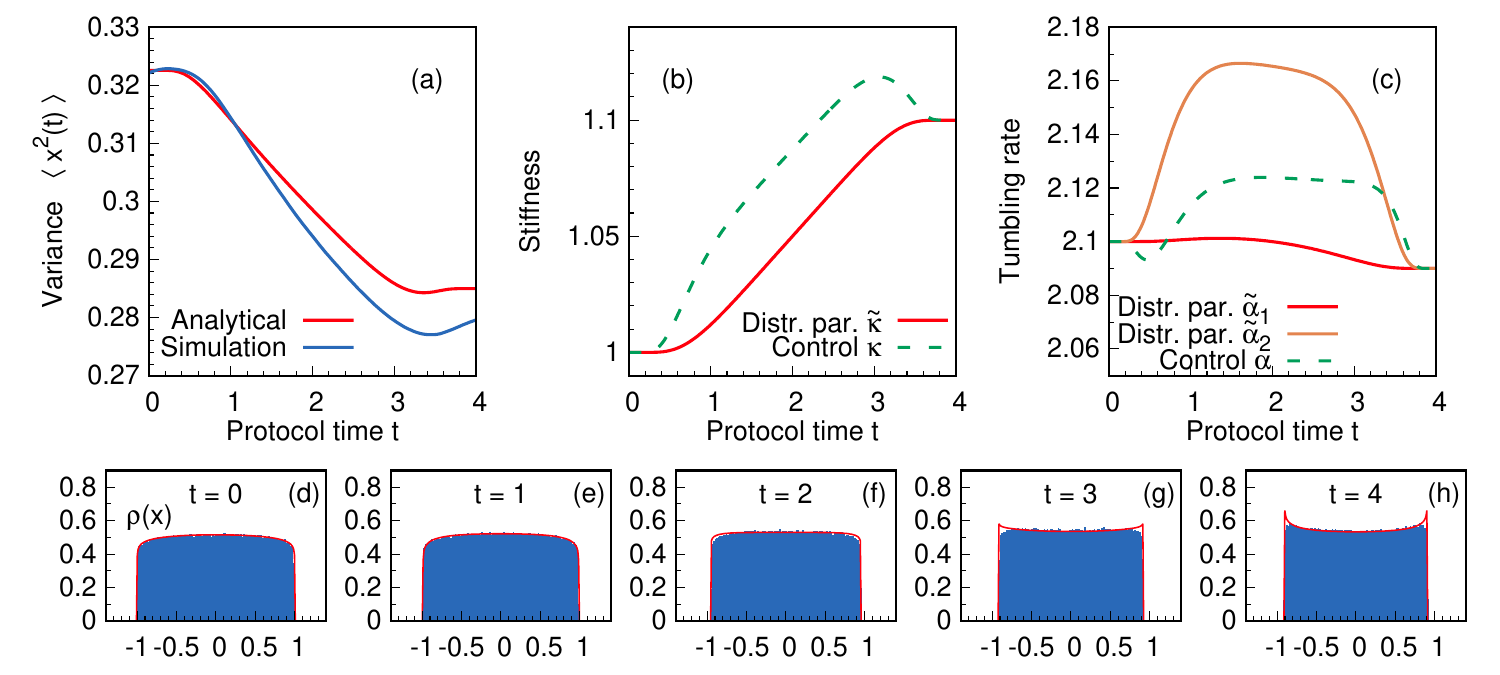}
    \caption{Nonoptimal shortcut in the \new{long-duration} regime. The protocol is found with the same procedure discussed in relation to Fig.~\ref{fig:nonopt1}. The boundary conditions are the same as in Fig.~\ref{fig:opt}.}
    \label{fig:nonopt}
\end{figure*}

 If the boundary conditions of the process are such that the approximations described in Sec.~\ref{sec:qs} can be expected to hold, also the optimal control problem  becomes addressable. We showed that, in such conditions, the instantaneous state of the system can be explicitly written, up to order $O(\e)$, as a function of $\tk$, $\b_1$ and their time derivatives. In particular, taking into account the functional form~\eqref{eq:rhopert} of the position distribution, the instantaneous variance reads
\begin{equation}
\label{eq:varhyper}
    \moy{x^2}=\frac{_1\mathbf{F}_1(3/2,5/2+\b_2,\b_1-\b_2)}{2\, \tk^2\, _1\mathbf{F}_1(1/2,3/2+\b_2,\b_1-\b_2)}
\end{equation}
where $_1\mathbf{F}_1(a,b,z)$ is the  Regularized Kummer confluent hypergeometric function of the variable $z$, with parameters $a$ and $b$~\cite{mathews2022physicist}. By substituting Eq.~\eqref{eq:closure2}, and expanding up to the first perturbative order in $\e$, one eliminates the dependence on $\b_2$, introducing in turn a dependence on $\b_1'$. We use again the prime symbol to refer to derivatives with respect to the rescaled time $s$ introduced in Section~\ref{sec:qs}.
 
In the light of the above, the Lagrangian~\eqref{eq:lag} can be also written as a function of $\tk$, $\b_1$ and their time derivatives, up to $O(\e^2)$:
\begin{equation}
    \label{eq:lagqs}
\mathcal{L}=\frac{\e \tk^2\tk'-\e^2 [(\tk')^2-\tk \tk'']}{\tk^4 (3+2\b_1)}+\e^2 \frac{\tk'}{\tk^3}\mathcal{F}(\b_1, \b_1')+O(\e^3)\,.   
\end{equation}
The explicit expression of the function $\mathcal{F}(\b_1, \b_1')$ is quite involved and not very informative. 
Let us notice that the Lagrangian is specified up to $O(\e^2)$. Since the time interval $t_f$ is $O(\e^{-1})$, this implies that we are neglecting corrections of order $\e^2$ for the total cost (i.e., the average work).

By inserting $\mathcal{L}$ in the Euler-Lagrange equations
    \begin{subequations}
    \label{eq:eulag}
    \begin{eqnarray}
        \frac{\partial \mathcal{L}}{\partial \b_1}- \frac{d}{ds} \frac{\partial \mathcal{L}}{\partial \b_1'} &=&0\\
        \frac{\partial \mathcal{L}}{\partial \tk} - \frac{d}{ds} \frac{\partial \mathcal{L}}{\partial \tk'} + \frac{d^2}{ds^2} \frac{\partial \mathcal{L}}{\partial \tk''} &=&0
    \end{eqnarray}
    \end{subequations}
one obtains a fourth-order differential system. This means that we need to fix four scalar quantities, and we can use this freedom to match the boundary conditions~\eqref{eq:boundcond}: in particular, we can specify the values of $\tk(0)$, $\tk(t_f)$, $\b_1(0)$ and $\b_1(t_f)$. The boundary value problem can be then solved numerically by means of ``shooting'' algorithms.

A comparison between an optimal and a non-optimal protocol with the same boundary conditions can be appreciated by looking at Figures~\ref{fig:opt} and~\ref{fig:nonopt}. As it is typical for optimal control problems~\cite{Schmiedl07prl}, the best strategy for the stiffness dynamics is characterized by a sudden change of the control at the initial time, enforcing the desired transformation on the distribution (in the considered case, a compression: see panel~\ref{fig:opt}(b), also compared to panel~\ref{fig:nonopt}(b)). The behaviour of $\a$ is less transparent: we will provide some considerations in Section~\ref{sec:act}.

Let us remark that in order to completely determine the distribution of the system~\eqref{eq:rhopert} at order $\e$ one would also need to specify the value of $\b_2$ at order $O(\e)$, or equivalently $\b_1'$ at order $O(1)$, at the end times. 
This means that the boundary states are matched, in principle, with an error of order $O(\e)$. This explains why, in Fig.~\ref{fig:opt}(c), the boundary conditions on $\widetilde{\a}_2$ (which is related to $\b_2$, see Eq.~\eqref{eq:a2}) do not coincide with those for $\widetilde{\a}_1$, as it would be expected for stationary states. The errors on the \old{end} \newAM{boundary} states \new{(initial and final)} also reflect on the variance plotted in Fig.~\ref{fig:opt}(a), to be compared with Fig.~\ref{fig:nonopt}(a).

\subsection{Dependence on the \blue{protocol time $t_f$}}

\begin{figure*}
    \centering
    \includegraphics[width=.8\linewidth]{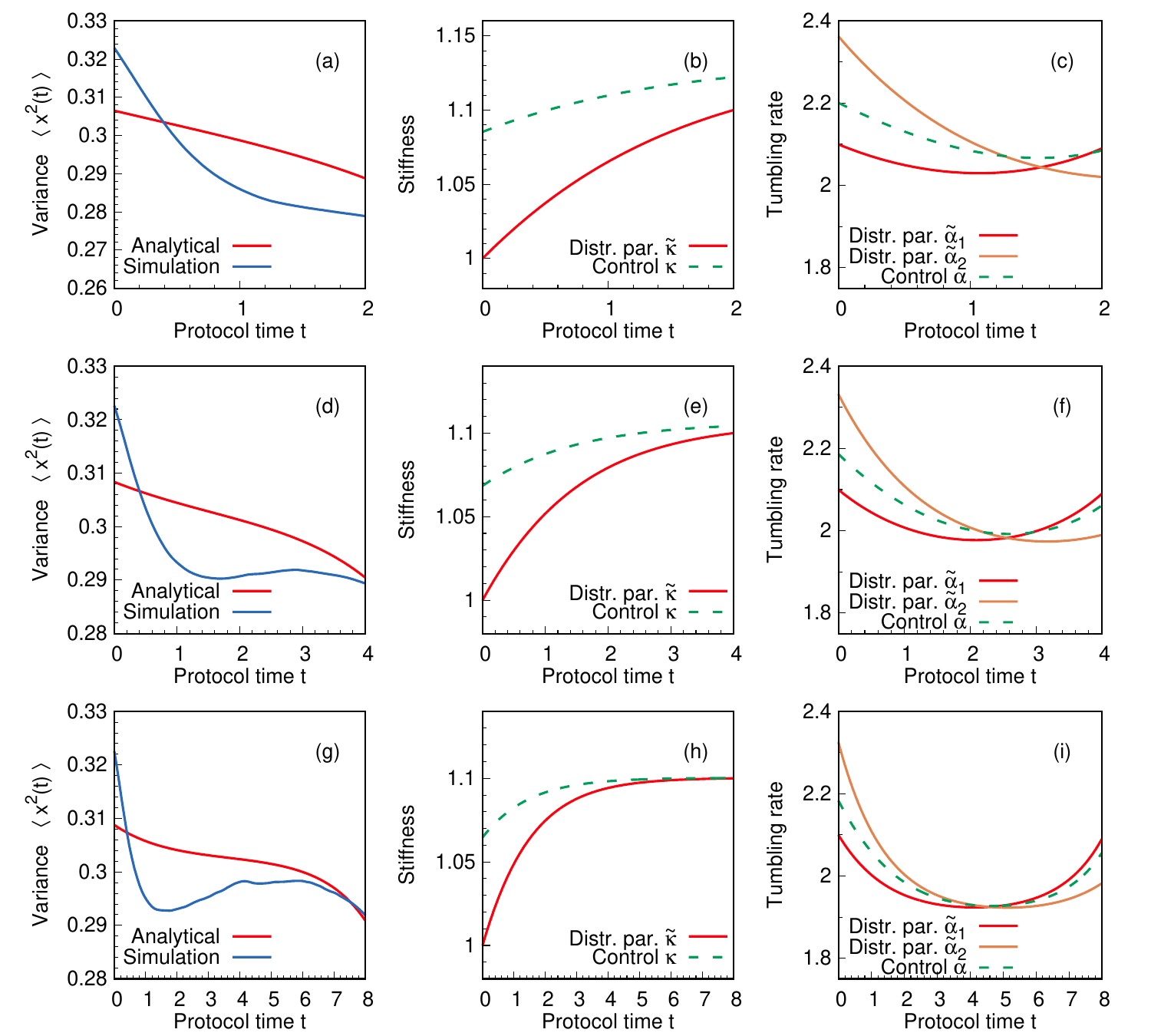}
    \caption{Optimal shortcuts in the \new{long-duration} regime, for increasing values of the \blue{protocol time $t_f$}. First, second and third row refer to the  $t_f=2$, $t_f=4$ and $t_f=8$ case, respectively. Boundary conditions and color code as in Fig.~\ref{fig:opt}.}
    \label{fig:tf}
\end{figure*}

\begin{figure}
    \centering
    \includegraphics[width=.99\linewidth]{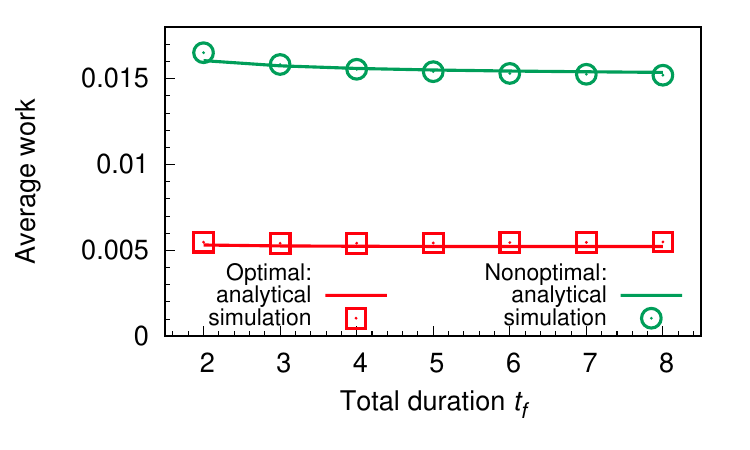}
    \caption{\new{Average work in the long-duration regime, as a function of the \blue{protocol time $t_f$}. The optimal case is represented in red: the continuous line shows the analytical solution, while points represent the results of numerical simulations. We also show the nonoptimal case, in green. All parameters are as in Figs.~\ref{fig:opt} and~\ref{fig:nonopt}.}}
    \label{fig:works}
\end{figure}

The derivation of the optimal protocol shown in Fig.~\ref{fig:opt} hinges on the slowness of the considered class of processes. It is therefore reasonable to expect that for increasing values of $t_f$ (i.e., decreasing values of $\e$), the provided approximation gets better and better. 

The performance of this approximation is shown in Fig.~\ref{fig:tf}, where the optimal control problem is solved for different values of $t_f$, in the presence of the same boundary conditions. We observe in particular that the agreement between the analytical calculation for the position variance and the behavior actually observed in numerical simulations improves as the \blue{protocol time $t_f$} increases (panels~\ref{fig:tf}(a),~\ref{fig:tf}(d) and~\ref{fig:tf}(g)).

\newAM{The optimization performance is shown in Fig.~\ref{fig:works}, where the work needed to perform the shortcut described in Fig.~\ref{fig:nonopt} is compared to the work during the optimal shortcut in Fig.~\ref{fig:opt}. The analytical results converge with the numerics already at relatively small values of $t_f$, with a slight deviation at small $t_f$ for the non-optimal protocol.}

\subsection{Optimal controls and activity}
\label{sec:act}
In this section we analyze in some more detail the specific features of the optimal protocols we found, arising from the presence of activity in the system.
It is important to keep in mind the caveat discussed above about the limitations of our procedure: specifically, the boundary conditions at the end times being fixed with an error $O(\e)$ implies that, strictly speaking, our results only hold true in the infinite $t_f$ limit. Nonetheless, we expect the qualitative features found here to be unaffected by this drawback.

\begin{figure*}
    \centering
    \includegraphics[width=.8\linewidth]{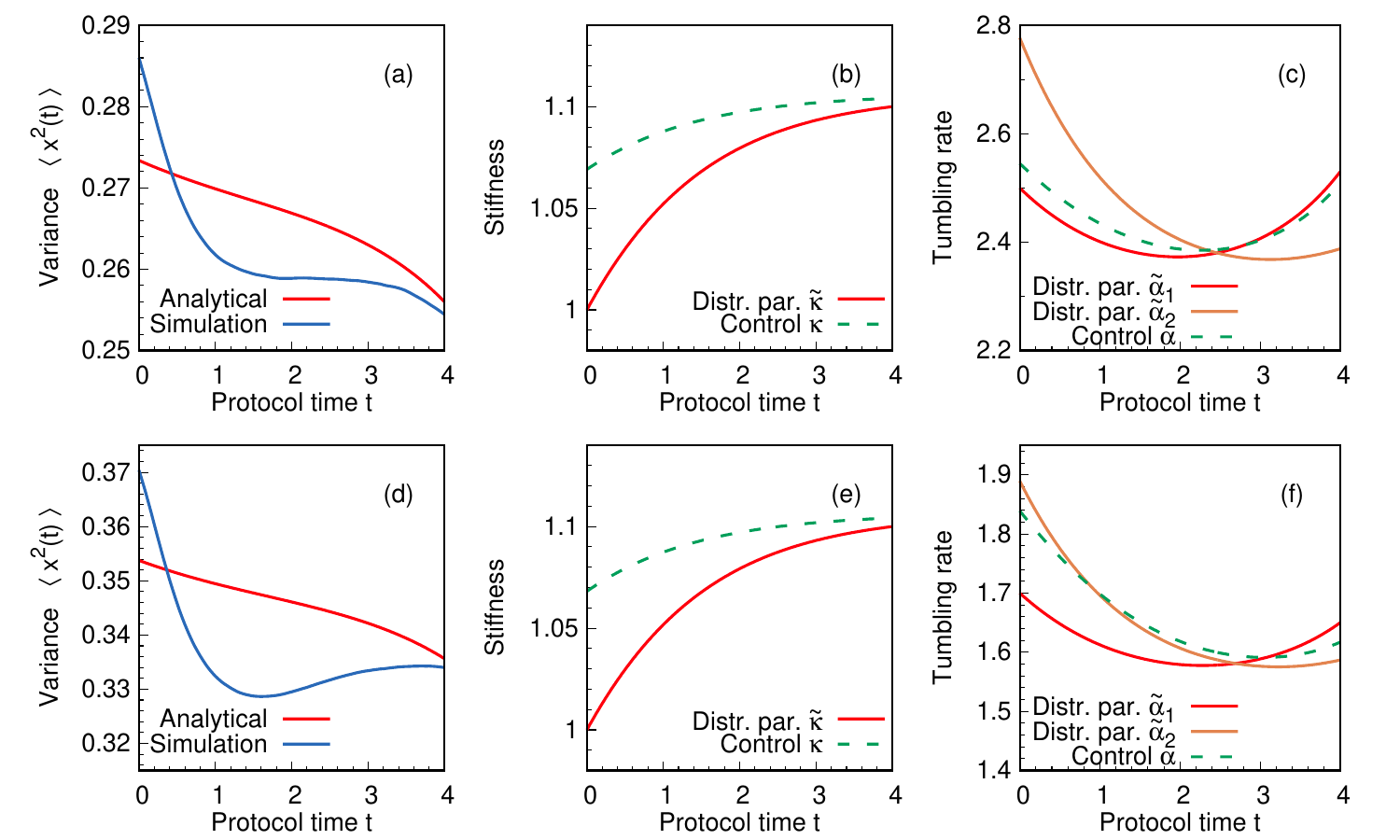}
    \caption{Optimal shortcut in the \new{long-duration} regime, in a passive-like and active-like situation. The first row shows the same analysis as in Fig.~\ref{fig:opt}, with $\b_i=0.25$, $\b_f=0.15$: both the initial and the final state are ``passive-like'' distributions. The second row shows instead the case $\b_i=-0.15$, $\b_f=-0.25$, where both end states are ``active-like''. }
    \label{fig:beta}
\end{figure*}

\begin{figure*}
    \centering
    \includegraphics[width=.8\linewidth]{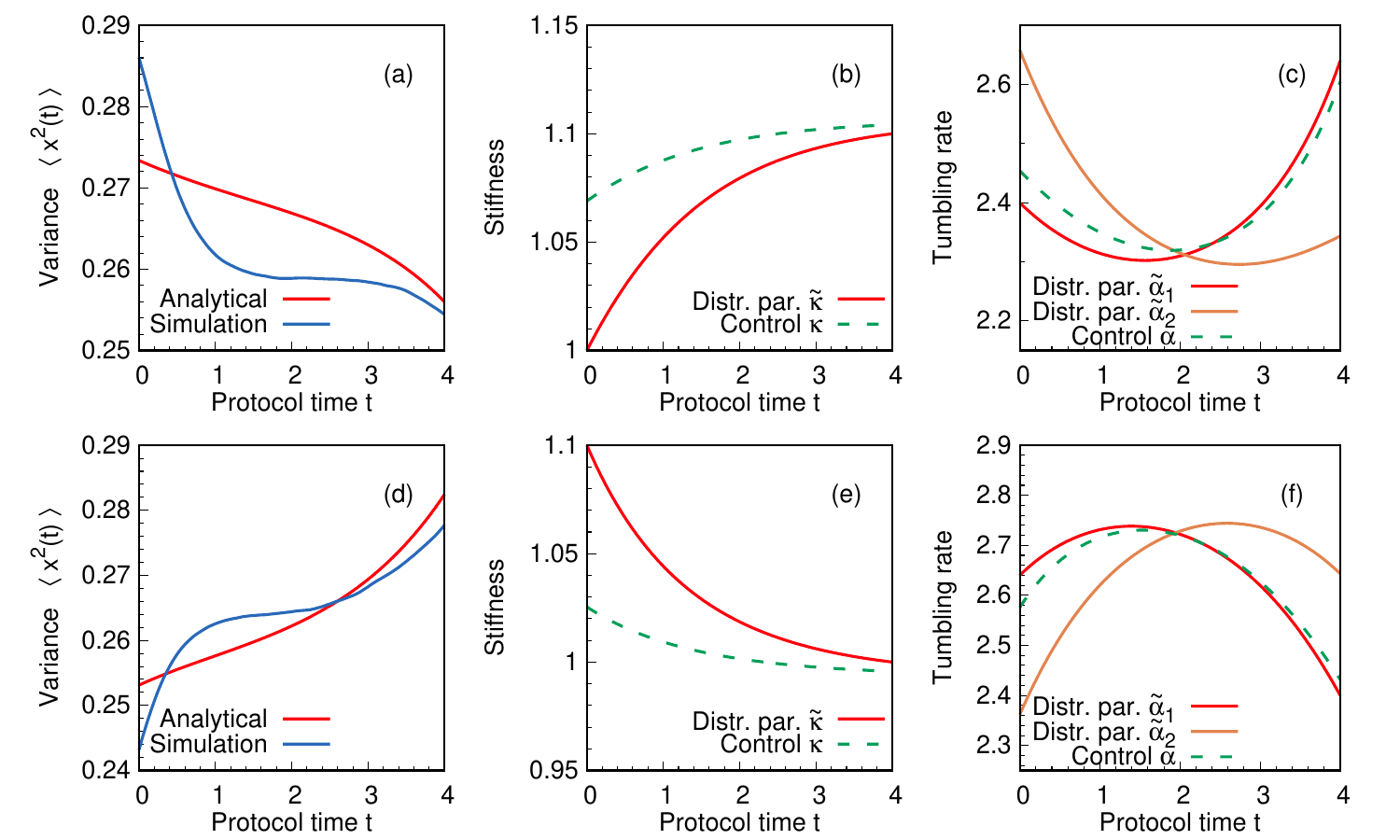}
    \caption{Optimal shortcut in the \new{long-duration} regime: compression vs decompression. With the same conventions used in the previous figures, the first row represents an optimal compression from $\tk=1$ to $\tk=1.1$, while the second row shows the inverse decompression from $\tk=1.1$ to $\tk=1$. In both cases $\b_i=\b_f=0.2$.}
    \label{fig:inverse}
\end{figure*}

In Fig.~\ref{fig:beta} we repeat the analysis shown in Fig.~\ref{fig:opt} for different boundary conditions on $\b_1$. We consider two cases: in the first one (panels~\ref{fig:beta}(a)-(c))  both the initial and the final state are ``passive-like'' (i.e., $\b_1>0$); in the second one (panels~\ref{fig:beta}(d)-(f)), they are both ``active-like'' ($\b_1<0$). 

The control protocol for the stiffness $\k(t)$ does not appear to depend much on the level of activity of the system. An efficient compression is always achieved by a sudden intensification of the confinement force, followed by a more gentle relaxation to the target state.

The interpretation of the control protocol for $\a(t)$ is less immediate. As a preliminary remark, we observe that when the system is in a passive-like condition, the parameter that responds more promptly to the control is $\widetilde{\a}_1$, while for higher levels of the activity the quantity that appears to be more directly affected is $\widetilde{\a}_2$. We suspect this behavior to be related to the 
roles of $\b_2$ and $\widetilde{\a}_2$ in our parametrization, as they encode  the non-Gaussian contributions to the distribution (i.e. those responsible for active features): see Eq.~\eqref{eq:rhopert}. Following this line of thought, to gain some intuition we restrict to a situation with $\b_1(0)=\b_1(t_f)>0$ and we focus on the behaviour of $\a$ and $\widetilde{\a}_1$. The results of the computation are shown in Fig.~\ref{fig:inverse}, where an optimal compression and an optimal decompression are compared.

In panel~\ref{fig:inverse}(c) it can be seen that the optimal compression relies on a sudden increase of $\a$ with respect to the boundary value of $\widetilde{\a}_1$ at the initial time, followed by a smooth protocol that gradually leads the system to the target distribution.
In other words, the less energy-consuming method to compress a system of active particles is to first... make them less active. This can be understood qualitatively by considering the particles located at the boundaries of the distribution, persistently oriented outwards, which are the signature of confined active systems. These particles push against the confining force, providing a positive contribution to the mechanical work that one aims to minimize. It is therefore sensible to start the protocol by decreasing their level of activity (i.e., by increasing $\a$).
A specular situation is encountered when one wants to perform an optimal decompression, passing from a higher to a lower value of $\tk$. In that case, as shown in panel~\ref{fig:inverse}(f), the best strategy is to start with an abrupt increase of the activity of the particles, exploiting in this way their persistent outward motion, which provides a negative contribution to the work.

\section{Conclusions}
\label{sec:conc}

In this paper we have studied control and optimal transport problems for a popular model of self-propelled particles. The RTP dynamics captures the most distinctive features of active matter, but its mathematical description is simple enough to make analytical calculations possible, at least to some extent.

First, \red{we have provided a constructive framework to design finite-time control protocols in active matter systems, based on a reformulation of the dynamics in terms of a hierarchy of observables}. Assuming that the strength of the confinement and the persistence time characterizing the dynamics can be controlled in time, we have proposed a recursive method to find shortcuts between stationary states. 
In the present paper we have studied and implemented its long-duration limit, which can be addressed analytically. 

In the same regime, we have identified control protocols minimizing the mechanical work operated on the system. As often observed in this kind of problems, the protocol starts with abrupt changes in the controls. In particular, we notice that for compression (expansion) transitions, the optimal protocol involves a sudden increase (decrease) of the confinement strength, even beyond the target one. This is in line with many results available in the literature. The behaviour of the parameter controlling the activity is, in principle, less obvious: we observe that, for compressions, it is more efficient to start the protocol by reducing the persistence time of the particles (making them ``less active''), while for expansions the opposite strategy is the best. This can be understood by recalling that persistent particles tend to move outwards, hence providing positive (negative) contribution to the work during compressions (expansions). We expect therefore this qualitative behaviour to be fairly general, and to be encountered in other kind of self-propelling particles.
\blue{We expect that the method proposed in this paper can be easily generalized to other models of self-propelled active matter, for instance to Active Brownian Particles or RTP in the presence of thermal noise: for these systems exact shortcuts are also available~\cite{Baldovin23prl,Frim23arxiv}, which could be used as benchmarks for the theory. We also expect the method} to be applicable to single-particle active systems characterized by qualitatively different mechanisms, e.g. pulsating units~\cite{Zhang2023,Manacorda2025}. \blue{Of course, our method requires the knowledge of the stationary distribution (at least in the form of a series expansion): extending it to cases with non-harmonic potentials is therefore a challenging task, as it happens also in the passive case~\cite{sanders2025minimal}. 
}

\blue{Our results also establish a direct connection with experimental strategies for controlling active matter, where external fields such as light are used to modulate particle motility~\cite{Buttinoni12jpcm,Vizsnyiczai17natcomm} and shape stationary density profiles~\cite{Frangipane18elife}. Our framework provides a systematic route to design time-dependent protocols achieving such target states with minimal dissipation. A natural extension of this approach concerns the design of optimal transformations between different density profiles. In particular, light-controlled bacterial systems offer the possibility of dynamically tuning activity in space and time, suggesting that transitions between nonequilibrium steady states could be experimentally optimized.}


In principle one could improve our results by expanding the Lagrangian function at the next perturbative order, or by introducing a multi-scale perturbative scheme. Both options are challenging, but they would allow for more accurate results, especially at the boundaries. As discussed in Section~\ref{sec:optqs}, indeed, our method involves errors of $O(\e)$ in the matching of the boundary conditions. The ultimate goal of this line of investigation would be the design of optimal cycles with RTP particles, possibly leading to a deeper understanding of the thermodynamics of active matter.

\section*{Data availability}
The data that support the findings of this article are openly available~\cite{zenodo}.

\begin{acknowledgments}
    The authors thank L. Angelani, M. Paoluzzi and A. Puglisi for their critical revision of the manuscript. \blue{A.M. thanks É. Fodor for useful discussions.}
    M.B. was supported by ERC Advanced Grant RG.BIO (Contract No. 785932) \blue{and by the MIUR PRIN 2022 (project ``SNO-MINK” no. 2022KWTEB7) which is funded by the European Union  Next Generation EU, M4 C2 1.1  CUP B53C24006470006.}
    A.M. acknowledges financial support from Grant No. 2022HNW5YL MOCA funded by the Ministero dell’Università e della Ricerca PRIN2022 program. 
\end{acknowledgments}

\appendix

\section{Derivation of Eqs.~\eqref{eq:gc}}
\label{app:comp}

If one evaluates Eqs.~\eqref{eq:dyn1} and~\eqref{eq:dyn2} at the boundaries of the spatial domain (i.e., at $x=\pm \tk^{-1}$), the derivatives of Heaviside step functions appear. Recalling Eq.~\eqref{eq:gc3a}, one gets:
\begin{subequations}
\nonumber
    \begin{eqnarray}
        \nonumber\sbr{\dot{\tk}+\g(x) \tk - \k \tk}\,\delta(1- \tk^2 x^2)&=&0\\
       \nonumber \sbr{\dot{\tk}+\g(x) \tk - \k \tk}\,\delta(\g(x) \dot{\tk} x^2 + \tk - \k \g(x) \tk x^2)&=&0\,.
    \end{eqnarray}
\end{subequations}
As a consequence,
$$
\gamma(\pm \tk^{-1})=\tk\,,
$$
and Eq.~\eqref{eq:gc_f5} holds.

By plugging Eq.~\eqref{eq:gc3a}  in Eq.~\eqref{eq:dyn1} one has
\beq
\label{eq:gca}
\partial_t \ln \r(x) = -x\g '(x)  - \g(x) + \k  -x\cbr{\g(x)-\k} \partial_x \ln \r(x)
\eeq
hence, recalling \eqref{eq:gc5} and \eqref{eq:gc3},
\beq
\label{eq:gc6}
\begin{aligned}
&-\partial_t \ln \cbr{\frac{\mathcal{Z}}{\tk}}- \sum_{n=1}^\infty \frac{\dot{\b_n}\tk^{2n}}{n}x^{2n}-2 \sum_{n=1}^\infty \b_n \tk^{2n}x^{2n}\frac{\dot{\tk}}{\tk}=
\\&-\sum_{n=0}^{\infty}c_n x^{2n}\argp{2n+1}+2\sum_{j=0}^{\infty}\sum_{l=0}^{\infty}c_j \b_l \tk^{2l}x^{2(l+j)}\\
&+\k-2\k\sum_{n=1}^{\infty}\b_n\tk^{2n}x^{2n}\,.    
\end{aligned}
\eeq
We can examine separately the different powers of $x$. At order $1$ we get
\beq
\label{eq:gc7a}
\partial_t \ln \cbr{\frac{\mathcal{Z}}{\tk}}=c_0-\k\,.
\eeq
Because of Eq.~\eqref{eq:gc_f5}, one immediately has Eq.~\eqref{eq:gc_f1}.
Higher powers of $x$ lead to
\beq
\label{eq:gc9}
\nonumber
c_n=\frac{2}{2n+1}\argp{\sum_{j=0}^{n-1}c_j\b_{n-j}\tk^{2\argp{n-j}} +\frac{\dot{\b_{n}}\tk^{2n}}{2n}-\b_n\tk^{2n+1}}\,,
\eeq
valid for $n\ge1$, where we have made use of Eq.~\eqref{eq:gc_f5}.
This equation provides a recursive expression for $c_n$, assuming that the $\arga{\b_j}_{j=1,..,n}$ are known. By introducing the time-dependent variables $\{\f_n\}_{n\ge1} $ as in Eq.~\eqref{eq:phi}, we readily get Eq.~\eqref{eq:gc_f4}.

Let us now consider Eq.~\eqref{eq:dyn2}. By inserting~\eqref{eq:gc3} one gets
\beq
\nonumber
x\partial_t \g  + x\g  \partial_t \ln \r = (\k x^2 \g-1) \partial_x \ln \r + 2 \k \g x + \k \g' x^2   - \a \g x\,,
\eeq
where we have dropped the explicit dependence on $x$. Using \eqref{eq:gca}, one obtains
\beq
\label{eq:gcb}
x\partial_t \g  - \g \g' x^2 - \g^2 x = (\g^2 x^2-1) \partial_x \ln \r +  \k \g x + \k \g' x^2 - \a \g x\,.
\eeq
We can now expand Eq.~\eqref{eq:gcb} using again Eqs.~\eqref{eq:gc5} and \eqref{eq:gc3}.
The coefficients of $x$ at the first order give
Eq.~\eqref{eq:gc_f3},
while the inspection of higher powers leads to
\begin{widetext}
\beq
\label{eq:gc12}
\dot{c}_n-\sum_{j=0}^{n}c_j c_{n-j}\argp{2n-2j+1}+2\sum_{j=0}^{n-1}\sum_{l=0}^{n-j-1}c_j c_l \b_{n-j-l}\tk^{2\argp{n-j-l}}=2\b_{n+1}\tk^{2\argp{n+1}}+\k c_n\argp{2n+1}-\a c_n\,, \quad n\ge1\,.
\eeq
Repeated use of Eqs.~\eqref{eq:gc9} and~\eqref{eq:gc_f5} yields
\beq
\label{eq:gc13}
\dot{c}_n=c_0 c_n +\argp{2n+1}\frac{\dot{\tk}}{\tk}c_n -2\b_n \tk^{2n+2}+ \frac{\dot{\b}_{n}\tk^{2n+1}}{n}+\sum_{j=0}^{n-1}\frac{c_j \dot{\b}_{n-j}\tk^{2\argp{n-j}}  }{n-j}+2\b_{n+1}\tk^{2\argp{n+1}}-\a c_n\,.
\eeq
By recalling Eq.~\eqref{eq:phi} one has
\beq
\label{eq:gc16}
\dot{c}_n=c_0 c_n + \argp{2n+1}\argp{\k-c_0} c_n+2 c_0\argp{\f_n-c_0 \b_n}\tk^{2 n}+\sum_{j=1}^{n-1}c_j \f_{n-j}\tk^{2\argp{n-j}}+2\b_{n+1}\tk^{2\argp{n+1}}-\a c_n\,.
\eeq
In order to get closed equations for the $\arga{\b_j}_{j\ge 1}$,  we take the time derivative of~\eqref{eq:gc_f2}
\beq
\nonumber
\argp{2n+1}\dot{c}_n = 2 \sum_{j=1}^{n-1}\dot{c}_{j} \b_{n-j}\tk^{2\argp{n-j}}+2\sum_{j=1}^{n-1}c_j \dot{\b}_{n-j}\tk^{2\argp{n-j}}+4\sum_{j=1}^{n-1}c_j\b_{n-j}\argp{n-j}\tk^{2\argp{n-j}}\frac{\dot{\tk}}{\tk}+2n\tk^{2n}\frac{\dot{\tk}}{\tk}\f_n+\tk^{2n}\dot{\f}_n
\eeq
 and we plug~\eqref{eq:gc16} in it, so to get rid of the $\dot{c}_j$ terms. The algebra is lengthy but straightforward: one must use Eq.~\eqref{eq:phi}, Eq.~\eqref{eq:gc_f2} and the identity
 \beq
\label{eq:gc18}
2\sum_{j=1}^{n-1}\sum_{l=1}^{j-1}\f_{j-l}c_l \b_{n-j}\tk^{2\argp{n-l}}=\sum_{j=1}^{n-1}\f_{n-j}\argp{2j+1}c_j \tk^{2\argp{n-j}}-\tk^{2n} \sum_{j=1}^{n-1}\f_j \f_{n-j}\,.
\eeq
The final result is Eq.~\eqref{eq:gc_f4}.

\end{widetext}

\section{A recursive strategy}
\label{sec:rec}

For the sake of definiteness, let us consider the case in which both the initial and the target state are stationary. The boundary conditions in our notation read
\begin{subequations}
\label{eq:boundcond}
\begin{eqnarray}
\tk(0)&=&\tk_i\\
\b_n(0)&=&\b_i \quad \forall n \ge 1\\
\tk(t_f)&=&\tk_f\\
\b_n(t_f)&=&\b_f \quad \forall n \ge 1\,.
\end{eqnarray}
\end{subequations}
As a first step, one should identify a pair of functions $\b_1(t)$ and $\tk(t)$ compatible with these constraints.  The requirements on $\{\b_n\}_{n\ge2}$ are automatically fulfilled if
\begin{subequations}
    \begin{eqnarray}
       \label{eq:simub} \b_1(t)&=&\b_i+\argp{\b_f-\b_i}g(t)\\
       \label{eq:simuk} \tk(t)&=&\tk_i+\argp{\tk_f-\tk_i}h(t)\,,
    \end{eqnarray}
\end{subequations}
where the functions $g(t)$, $h(t)$ can be arbitrarily chosen among those verifying
\blue{
$$
g(0) = h(0) = 0\,,\quad g(t_f ) = h(t_f ) = 1
$$
and}
$$
\partial_t^n g(0)=\partial_t^n g(t_f)=\partial_t^n h(0)=\partial_t^n h(t_f)=0 \,, \quad \forall n\ge 1\,.
$$
\blue{The last condition is needed in order for Eqs.~\eqref{eq:gc} to represent a stationary state at $t=0$, $t=t_f$.}

A recursive strategy to obtain shortcuts for our system may proceed as follows:
\begin{enumerate}
    \item make an initial guess for the functional form of $w(t)$, e.g. $w_{\text{guess}}(t):=\partial_t \ln \mathcal{Z}_s(\b_1)$ (as if the state was stationary at every time, see Eq.~\eqref{eq:rhos-harmonic-ZZ});
    \item \label{point} compute $\{\b_n(t)\}_{2\le n\le M}$ recursively by means of Eqs.~\eqref{eq:gc}, where the cutoff $M$ depends on the required accuracy;
    \item apply definition~\eqref{eq:w}, \new{assuming $\b_n=\b_M$ for every $n>M$,} to make a new guess for $w(t)$:
    $$
    w_{\text{guess}}(t):= \partial_t \ln \mathcal{Z}(\{\b_n\})\,;
    $$
    \item iterate from point~\ref{point} until convergence.
\end{enumerate}
Establishing convergence conditions for the above scheme might be challenging. Even disregarding this issue, the implementation of the method involves a series of technical subtleties, such as the amplification of numerical errors due to the numerical computation of the time derivatives in Eq.~\eqref{eq:gc_f4}, and the nontrivial choice of the cutoff $M$. We leave therefore the practical implementation of this strategy for future studies.


\section{Details on the numerical simulations}
\label{app:simul}

The analytical results presented in this paper are always compared to the outcome of numerical simulations. These are obtained by applying the analytically computed protocol to a system of independent particles evolving according to Eq.~\eqref{eq:micro1}.

The initial state (which is always stationary in our examples) is prepared by extracting randomly the initial position and orientation of a particle, and then evolving it by an event-driven dynamics with $\k=\tk_i$ and $\a=2 \tk_i(1+\b_i)$. To ensure convergence to a stationary state, the event-driven dynamics is run for a time $t_*$ much longer than the typical relaxation times of the system (in the considered cases, we always impose $t_*=50\, t_f$).

An event is represented by a change of the particle's orientation $\s$. Denoting by $\{t_j\}_{j\ge1}$ the times at which such events occur, with $t_0=-t_*$, the interval between consecutive times is randomly extracted from the exponential distribution 
$$
(t_{j+1}-t_{j}) \sim \a e^{- \a (t_{j+1}-t_{j})}\,.
$$
Once $\s(t)$ is known for $-t_*\le t \le 0$, the dynamics of the particle is completely determined in such time interval, and its position at time $t=0$ can be easily computed from Eq.~\eqref{eq:micro1}.

The dynamics occurring in the time interval $0\le t \le t_f$ is subject to changing values of the controls, and the event-driven strategy cannot be used. We switch therefore to an integration scheme with time steps of fixed  duration $\D t$ much smaller than the characteristic times of the dynamics (in our examples, $\D t=0.01$). We make the approximation that the control parameters $\a(t)$ and $\tk(t)$ can be considered constant within such time steps. In each of them, the position can be therefore updated according to Eq.~\eqref{eq:micro1}, while the sign of $\sigma(t)$ is flipped with a probability $\a(t) \D t/2$.

The whole procedure is repeated $N \gg 1$ times (in our examples, $N=10^6$) to have reliable statistics for the computation of histograms and averages.

\bibliography{biblio}

\end{document}